\documentclass[12pt]{article}
\def\be{\begin{equation}}
\def\ee{\end{equation}}
\def\bea{\begin{eqnarray}}
\def\eea{\end{eqnarray}}
\usepackage{graphicx}

\catcode`\@=11
\def\lsim{\mathrel{\mathpalette\@versim<}}
\def\gsim{\mathrel{\mathpalette\@versim>}}
\def\@versim#1#2{\vcenter{\offinterlineskip
\ialign{$\m@th#1\hfil##\hfil$\crcr#2\crcr\sim\crcr } }}
\catcode`\@=12

\catcode`@=12
\begin{document}
\thispagestyle{empty}
\begin{flushright}
UCRHEP-T517\\
April 2012\
\end{flushright}
\vspace{0.1in}
\begin{center}
{\Large \bf CP Phases of Neutrino Mixing in a Supersymmetric\\
$B-L$ Gauge Model with $T_7$ Lepton Flavor Symmetry\\}
\vspace{0.3in}
{\bf Hajime Ishimori$^1$, Shaaban Khalil$^{2,3}$
and Ernest Ma$^{4,5,6}$\\}
\vspace{0.2in}
{\sl $^1$ Department of Physics, Kyoto University,
Kyoto 606-8502, Japan\\}
\vspace{0.1in}
{\sl $^2$ Centre for Theoretical Physics, Zewail City of Science and Technology,\\
Sheikh Zayed, 6 October City, 12588, Giza, Egypt\\} \vspace{0.1in}
{\sl $^3$ Department of Mathematics, Ain Shams University,\\
Faculty of Science, Cairo 11566, Egypt\\}
\vspace{0.1in}
{\sl $^4$ Department of Physics and Astronomy, University of
California,\\
Riverside, California 92521, USA\\}
\vspace{0.1in}
{\sl $^5$  Institute for Advanced Study, Hong Kong University of Science
and Technology,\\
Hong Kong, China\\}
{\sl $^6$  Kavli Institute for the Physics and Mathematics of the Universe,\\ 
University of Tokyo, Kashiwa 277-8583, Japan\\}
\end{center}
\vspace{0.3in}
\begin{abstract}\
In a recently proposed renormalizable model of neutrino mixing using the
non-Abelian discrete symmetry $T_7$ in the context of a supersymmetric
extension of the Standard Model with gauged $U(1)_{B-L}$, a correlation
was obtained between $\theta_{13}$ and $\theta_{23}$ in the case where all
four parameters are real.  Here we consider one parameter to be complex, thus
allowing for one Dirac CP phase $\delta_{CP}$ and two Majorana CP phases
$\alpha_{1,2}$.  We find a slight modification to this correlation as a 
function of $\delta_{CP}$.  For a given set of input values of 
$\Delta m^2_{21}$, $\Delta m^2_{32}$, $\theta_{12}$, and $\theta_{13}$, 
we obtain $\sin^2 2 \theta_{23}$ and $m_{ee}$ (the effective Majorana 
neutrino mass in neutrinoless double beta decay) as functions of $\tan 
\delta_{CP}$.  We find that the structure of this model always yields 
small $|\tan \delta_{CP}|$.
\end{abstract}

\newpage
\baselineskip 24pt

The most general $3 \times 3$ Majorana neutrino mass matrix has six complex
entries, i.e. twelve parameters.  Three are overall phases of the mass
eigenstates which are unobservable.  The nine others are three masses,
three mixing angles, and three phases: one Dirac phase $\delta_{CP}$, i.e.
the analog of the one complex phase of the $3 \times 3$ quark mixing matrix,
and two relative Majorana phases $\alpha_{1,2}$ for two of the three mass
eigenstates. The existence of nonzero $\delta_{CP}$ or $\alpha_{1,2}$ means
that $CP$ conservation is violated.  It is one of most important issues of
neutrino physics yet to be explored experimentally.

The application of the non-Abelian discrete symetry $A_4$~\cite{mr01}
(and others) to neutrino mixing has been successful in explaining
tribimaximal mixing, i.e. $\sin^2 \theta_{12} = 1/3$,
$\sin^2 \theta_{23} = 1/2$, and $\theta_{13} = 0$.  In particular, a
generic three-parameter $A_4$ model~\cite{m04} predicts all of the above
with $\delta_{CP} = \alpha_{1,2} = 0$, leaving the three neutrino masses
arbitrary.  Recently, the first evidence that $\theta_{13} \neq 0$ has been
published~\cite{t2k11} by the T2K Collaboration, i.e.
\begin{equation}
0.03 (0.04) \leq \sin^2 2 \theta_{13} \leq 0.28 (0.34)
\end{equation}
for $\delta_{CP}=0$ and normal (inverted) hierarchy of neutrino masses.
Slightly different but similar ranges are obtained for nonzero values
of $\delta_{CP}$.  More recently, the Double Chooz Collaboration has also
reported~\cite{dc11} a measurement of
\begin{equation}
\sin^2 2 \theta_{13} = 0.086 \pm 0.041({\rm stat}) \pm 0.030({\rm syst}).
\end{equation}
Their best fit is obtained by minimizing its $\chi^2$ as a function of
$\delta_{CP}$.  However, all $\delta_{CP}$ values are allowed within one
standard deviation.  One month ago, the first precise measurement of
$\sin^2 2 \theta_{13}$ was announced by the Daya Bay
Collaboration~\cite{daya12}:
\begin{equation}
\sin^2 2 \theta_{13} = 0.092 \pm 0.016({\rm stat}) \pm 0.005({\rm syst}),
\end{equation}
based only on a rate analysis, resulting in a 5.2$\sigma$ effect.
It has been followed by the announcement this month of the RENO
Collaboration~\cite{reno12}:
\begin{equation}
\sin^2 2 \theta_{13} = 0.103 \pm 0.013({\rm stat}) \pm 0.011({\rm syst}),
\end{equation}
again based only on a rate analysis, resulting in a 6.3$\sigma$ effect.

To account for $\theta_{13} \neq 0$, the original $A_4$ proposal has to be
modified~\cite{mw11}.  Similarly, the original supersymmetric
$B-L$ gauge model with $T_7$ lepton flavor symmetry~\cite{ckmo11}
(which obtained tribimaximal mixing) has to be replaced as
well~\cite{ckmo11R}.  In that latter paper, it is shown that a neutrino mass
matrix with four parameters allow a nonzero $\theta_{13}$.  Assuming
that all four parameters are real, thus requiring two conditions
among the six observables, i.e. the three masses and three mixing
angles, the prediction
\begin{equation}
\sin^2 2 \theta_{23} \simeq 1 - {1 \over 2} \sin^2 2 \theta_{13}
\end{equation}
is obtained.  This scenario applies of course only to the case
$\delta_{CP} = \alpha_{1,2} = 0$.  Here we consider instead the
case where one parameter is complex.  We then have five real parameters 
to describe the three masses, the three mixing angles, and the three phases.  
Given five inputs, we should then be able to predict the other four parameters.

The tetrahedral group $A_4$ (12 elements) is the smallest group with a real
\underline{3} representation. The Frobenius group $T_7$ (21 elements) is
the smallest group with a pair of complex \underline{3} and \underline{3}$^*$
representations.  It is generated by
\begin{equation}
a = \pmatrix{\rho & 0 & 0 \cr 0 & \rho^2 & 0 \cr 0 & 0 & \rho^4}, ~~~
b = \pmatrix{0 & 1 & 0 \cr 0 & 0 & 1 \cr 1 & 0 & 0},
\end{equation}
where $\rho = \exp(2 \pi i /7)$, so that $a^7=1$, $b^3=1$, and $ab = ba^4$.
The character table of $T_7$ (with $\xi = -1/2 + i \sqrt{7}/2$) is given 
below.

\begin{table}[htb]
\centerline{\begin{tabular}{|c|c|c|c|c|c|c|c|}
\hline
class & $n$ & $h$ & $\chi_1$ & $\chi_{1'}$ & $\chi_{1''}$ & $\chi_3$ &
$\chi_{3^*}$ \\
\hline
$C_1$ & 1 & 1 & 1 & 1 & 1 & 3 & 3 \\
$C_2$ & 7 & 3 & 1 & $\omega$ & $\omega^2$ & 0 & 0 \\
$C_3$ & 7 & 3 & 1 & $\omega^2$ & $\omega$ & 0 & 0 \\
$C_4$ & 3 & 7 & 1 & 1 & 1 & $\xi$ & $\xi^*$ \\
$C_5$ & 3 & 7 & 1 & 1 & 1 & $\xi^*$ & $\xi$ \\
\hline
\end{tabular}}
\caption{Character table of $T_7$.}
\end{table}

The group multiplication rules of $T_7$ include
\begin{eqnarray}
\underline{3} \times \underline{3} &=& \underline{3}^* (23,31,12) +
\underline{3}^* (32,13,21) + \underline{3} (33,11,22), \\
\underline{3} \times \underline{3}^* &=& \underline{3} (2 1^*, 3 2^*, 1 3^*) +
\underline{3}^* (1 2^*, 2 3^*, 3 1^*) + \underline{1} (1 1^* + 2 2^* + 3 3^*)
\nonumber \\  &+& \underline{1}' (1 1^* + \omega 2 2^* + \omega^2 3 3^*) +
\underline{1}'' (1 1^* + \omega^2 2 2^* + \omega 3 3^*).
\end{eqnarray}
Note that $\underline{3} \times \underline{3} \times \underline{3}$ has two
invariants and $\underline{3} \times \underline{3} \times \underline{3}^*$
has one invariant.

We now follow Ref.~\cite{ckmo11R} in deriving the neutrino mass matrix.
Under $T_7$, let $L_i = (\nu,l)_i \sim \underline{3}$, $l^c_i \sim
\underline{1},\underline{1}',\underline{1}'',~i=1,2,3$, $\Phi_i =
(\phi^+,\phi^0)_i \sim \underline{3}$,
and ${\Phi'}_i = ({\phi'}^0,-{\phi'}^-)_i \sim \underline{{3^*}}$.
The Yukawa couplings $L_i l^c_j {\Phi'}_k$ generate the charged-lepton
mass matrix
\bea
M_l = \pmatrix{f_1 v'_1 & f_2 v'_1 & f_3 v'_1 \cr f_1 v'_2 & \omega^2 f_2 v'_2 &
\omega f_3 v'_2 \cr f_1 v'_3 & \omega f_2 v'_3 & \omega^2 f_3 v'_3}
= {1 \over \sqrt{3}} \pmatrix{1 & 1 & 1 \cr 1 & \omega^2 & \omega \cr 1
& \omega
& \omega^2} \pmatrix{f_1 & 0 & 0 \cr 0 & f_2 & 0 \cr 0 & 0 & f_3} ~v,
\eea
if $v'_1 = v'_2 = v'_3 = v'/\sqrt{3}$, as in the original $A_4$
proposal~\cite{mr01}.

Let $\nu^c_i \sim \underline{{3^*}}$, then the Yukawa couplings
$L_i \nu^c_j \Phi_k$ are allowed, with
\begin{equation}
M_D = f_D \pmatrix{0 & v_1 & 0 \cr 0 & 0 & v_2 \cr v_3 & 0 & 0} =
{f_D v \over \sqrt{3}} \pmatrix{0 & 1 & 0 \cr 0 & 0 & 1 \cr 1 & 0 & 0},
\end{equation}
for $v_1 = v_2 = v_3 = v/\sqrt{3}$ which is necessary for consistency
since $v'_1 = v'_2 = v'_3 = v'/\sqrt{3}$ has already been assumed for $M_l$.
Note that $\Phi$ and ${\Phi'}$ have $B-L=0$, and both are necessary because
of supersymmetry.  However, the analysis of neutrino mixing does not 
involve these extra supersymmetric partners.

Now add the neutral electroweak Higgs singlets $\chi_i \sim \underline{3}$ 
and $\eta_i \sim \underline{{3^*}}$, both with $B-L=-2$.  Then there are 
two Yukawa
invariants: $\nu^c_i \nu^c_j \chi_k$ and $\nu^c_i \nu^c_j \eta_k$ (which has
to be symmetric in $i,j$).  Note that $\chi_i^* \sim \underline{{3^*}}$
is not the same as $\eta_i \sim \underline{{3^*}}$ because they have
different $B-L$.  This means that both $B-L$ and the complexity of the
$\underline{3}$ and $\underline{{3^*}}$ representations in $T_7$ are
required for this scenario. The heavy Majorana mass matrix for $\nu^c$ is then
\bea
M_{\nu^c} = h \pmatrix{u_2 & 0 & 0 \cr 0 & u_3 & 0 \cr 0 & 0 & u_1} +
h' \pmatrix{0 & u'_3 & u'_2 \cr u'_3 & 0 & u'_1 \cr u'_2 & u'_1 & 0} =
\pmatrix{A & C & B \cr C & D & C \cr B & C & D},
\eea
where $A = h u_2$, $B = h' u'_2$, $C = h' u'_1 = h' u'_3$, and
$D = h u_1 = h u_3$ have been assumed.  This means that the residual
symmetry in the singlet Higgs sector is $Z_2$, whereas that in
the doublet Higgs sector is $Z_3$.  This choice allows nonzero $\theta_{13}$,
whereas the choice of Ref.~\cite{ckmo11} enforces $\theta_{13}=0$.

The seesaw neutrino mass matrix is now
\bea
M_\nu = - M_D M_{\nu^c}^{-1} M_D^T
 = {- f_D^2 v^2 \over 3~{\rm det}(M_{\nu^c})} \pmatrix{AD-B^2
& C(B-A) & C(B-D) \cr C(B-A) & AD-C^2 & C^2-BD \cr C(B-D) & C^2-BD & D^2-C^2},
\eea
where det$(M_{\nu^c}) = A(D^2-C^2) + 2BC^2 - D(B^2+C^2)$.  Redefining the
parameters $A,B,C,D$ to absorb the overall constant, we obtain the
following neutrino mass matrix in the tribimaximal basis:
\begin{equation}
{\cal M}_\nu^{(1,2,3)} = \pmatrix{D(A+D-2B)/2 & C(2B-A-D)/\sqrt{2} &
D(A-D)/2 \cr C(2B-A-D)/\sqrt{2} & AD-B^2 & C(D-A)/\sqrt{2} \cr
D(A-D)/2 & C(D-A)/\sqrt{2} & (AD+D^2+2BD-4C^2)/2}.
\end{equation}
This is obtained by first rotating with the $3 \times 3$ unitary matrix of
Eq.~(9), which converts it to the $(e, \mu, \tau)$ basis, then by Eq.~(14)
below.  Note that for $D=A$ and $C=0$, this matrix becomes diagonal:
$m_1 = A(A-B), m_2 = A^2-B^2, m_3 = A(A+B)$, which is the tribimaximal limit.
Normal hierarchy of neutrino masses is obtained if $B \simeq A$ and inverted
hierarchy is obtained if $B \simeq -2A$.

The neutrino mixing matrix $U$ has 4 parameters: $s_{12}, s_{23}, s_{13}$ and
$\delta_{CP}$~\cite{pdg10}.  We choose the convention $U_{\tau 1}, U_{\tau 2},
U_{e3}, U_{\mu 3} \to -U_{\tau 1}, -U_{\tau 2},
-U_{e3}, -U_{\mu 3}$ to conform with that of the tribimaximal mixing matrix
\begin{equation}
U_{TB} = \pmatrix{\sqrt{2/3} & 1/\sqrt{3} & 0 \cr -1/\sqrt{6} & 1/\sqrt{3} &
-1/\sqrt{2} \cr -1/\sqrt{6} & 1/\sqrt{3} & 1/\sqrt{2}}.
\end{equation}
then
\begin{equation}
{\cal M}_\nu^{(1,2,3)} = \pmatrix{m_1 & m_6 & m_4 \cr m_6 & m_2 & m_5 \cr
m_4 & m_5 & m_3} = U^T_{TB} U \pmatrix{e^{i\alpha_1} m'_1 & 0 & 0 \cr
0 & e^{i\alpha_2} m'_2 & 0 \cr 0 & 0 &  m'_3} U^T U_{TB},
\end{equation}
where $m'_{1,2,3}$ are the physical neutrino masses, with
\begin{eqnarray}
m'_2 &=& \sqrt{{m'_1}^2 + \Delta m^2_{21}}, \\
m'_3 &=& \sqrt{{m'_1}^2 + \Delta m^2_{21}/2 + \Delta m_{32}^2}~~{\rm (normal
~hierarchy)}, \\
m'_3 &=& \sqrt{{m'_1}^2 + \Delta m^2_{21}/2 - \Delta m_{32}^2}~~{\rm
(inverted~hierarchy)}.
\end{eqnarray}
If $U$ and $\alpha_{1,2}$ are known, then all $m_{1,2,3,4,5,6}$ are functions
only of $m'_1$.

In Ref.~\cite{ckmo11R}, the parameters $A,B,C,D$ are assumed to be real,
hence $\delta_{CP}$ and $\alpha_{1,2}$ are zero.  We now consider 
$C = E + iF$ to be complex.  Thus $m_{1,2,4}$ are real and $m_{3,5,6}$
are complex.  Since ${\cal M}_\nu^{(1,2,3)}$ is in the tribimaximal basis,
it can be diagonalized by an approximately diagonal unitary matrix.  To 
first order, let 
\begin{equation}
U_\epsilon = \pmatrix{1 & \epsilon_{12} & \epsilon_{13} \cr
-\epsilon_{12}^* & 1 & \epsilon_{23} \cr
-\epsilon_{13}^* & -\epsilon_{23}^* & 1},
\end{equation}
then using
\begin{equation}
U_\epsilon {\cal M}_\nu^{(1,2,3)} U_\epsilon^T = \pmatrix{e^{i \alpha'_1}
m'_1 & 0 & 0 \cr 0 & e^{i \alpha'_2} m'_2 & 0 \cr 0 & 0 & e^{i \alpha'_3} 
m'_3},
\end{equation}
we obtain $\epsilon_{12}$, $\epsilon_{13}$, $\epsilon_{23}$ and 
$\alpha'_{1,2,3}$ in terms of $A,B,D,E,F$.  Using the four measured values 
$\Delta m^2_{21}$, $\Delta m^2_{32}$, $s_{12}$, $s_{13}$, and varying 
$\delta_{CP}$, we then obtain $s_{23}$, the physical relative Majorana 
phases $\alpha_{1,2}$ in Eq.~(15), and the effective 
Majorana neutrino mass in neutrinoless double beta decay, i.e.
\begin{equation}
m_{ee} = |U_{e1}^2 e^{i \alpha_1} m'_1 + U_{e2}^2 e^{i \alpha_2}
m'_2 + U_{e3}^2 m'_3|.
\end{equation}
Because of the structure of Eq.~(13) from the $T_7$ symmetry, even though
the phase of the complex parameter $C$ may be large, i.e. $F/E$ large,
$\tan \delta_{CP}$ cannot be too large, because in the limit $C=0$,
there can be no $CP$ violation, so any $CP$ violating effect has to
be proportional to $F/D$ where $D$ sets the neutrino mass scale.
This is typically less than one because $C \neq 0$ measures the
deviation of $\tan^2 \theta_{12}$ from the tribimaximal limit of 1/2.

The unitary matrix $U' = U_{TB} U^T_\epsilon$ has entries
\begin{eqnarray}
&& U'_{e1} = \sqrt{2 \over 3} + \sqrt{1 \over 3} \epsilon_{12}, ~~~
U'_{e2} = \sqrt{1 \over 3} - \sqrt{2 \over 3} \epsilon_{12}^*, ~~~
U'_{e3} = -\sqrt{2 \over 3} \epsilon_{13}^* - \sqrt{1 \over 3}
\epsilon_{23}^*, \\
&& U'_{\mu 3} = -{1 \over \sqrt{2}} + {\epsilon_{13}^* \over \sqrt{6}}
- {\epsilon_{23}^* \over \sqrt{3}}, ~~~
U'_{\tau 3} = {1 \over \sqrt{2}} + {\epsilon_{13}^* \over \sqrt{6}}
- {\epsilon_{23}^* \over \sqrt{3}}.
\end{eqnarray}
To obtain $U$, we rotate the phases of the $\mu$ and $\tau$ rows so that
$U'_{\mu 3} e^{-i \alpha'_3/2}$ is real and negative, and 
$U'_{\tau 3} e^{-i \alpha_3/2}$ is real and positive.
These phases are absorbed by the $\mu$ and $\tau$ leptons and are
unobservable.  We then rotate the $\nu_{1,2}$ columns so that 
$U'_{e1} e^{-i \alpha_3/2} = U_{e1} e^{i \alpha''_1/2}$ and 
$U'_{e2} e^{-i \alpha_3/2} = U_{e2} e^{i \alpha''_2/2}$, where 
$U_{e1}$ and $U_{e2}$ are real and positive.  
The physical relative Majorana phases of $\nu_{1,2}$ are then 
$\alpha_{1,2} = \alpha'_{1,2} + \alpha''_{1,2}$. 
We now extract the three angles as well as $\delta_{CP}$ as follows.
\begin{eqnarray}
&& \tan^2 \theta_{12} = \left| {U'_{e1} \over U'_{e2}} \right|^2 =
\left( {1 \over 2} \right) {(1- \sqrt{2} Re(\epsilon_{12}))^2 +
2 (Im(\epsilon_{12}))^2 \over (1+ Re(\epsilon_{12})/\sqrt{2})^2 +
 (Im(\epsilon_{12}))^2/2}, \\
&& \tan^2 \theta_{23} = \left| {U'_{\mu 3} \over U'_{\tau 3}} \right|^2 =
{(1- (Re(\epsilon_{13} - \sqrt{2} \epsilon_{23})/\sqrt{3})^2 +
(Im(\epsilon_{13} - \sqrt{2}\epsilon_{23}))^2/3 \over
(1+ (Re(\epsilon_{13} - \sqrt{2} \epsilon_{23})/\sqrt{3})^2 +
(Im(\epsilon_{13} - \sqrt{2}\epsilon_{23}))^2/3}, \\
&& \sin \theta_{13} e^{-i \delta_{CP}} = U'_{e3} e^{-i \alpha'_3/2}.
\end{eqnarray}

To see the approximate dependence of $U_\epsilon$ on the $T_7$ parameters 
$A,B,D,E,F$, we assume normal hierarchy and let
\begin{equation}
A = D + \delta_1, ~~~ B = D + \delta_2, ~~~ C = E + iF.
\end{equation}
Expanding in $\delta_{1,2}, E, F$ over $D$, we then have 
\begin{eqnarray}
&& m_1 = {D \over 2} (\delta_1 - 2 \delta_2), ~~~
m_2 = D (\delta_1 - 2 \delta_2) - \delta_2^2, ~~~
m_3 = 2 D^2 + {D \over 2}(\delta_1 + 2 \delta_2), \\
&& m'_1 = m_1 - {\delta_1^2 \over 8}, ~~~ m'_2 = m_2, ~~~ m'_3 = m_3, \\
&& m_4 = {D \over 2} \delta_1, ~~~ m_5 = - {\delta_1 \over \sqrt{2}}
(E + iF), ~~~ m_6 = - {E + iF \over \sqrt{2}} (\delta_1 - 2 \delta_2), \\
&& Re(\epsilon_{12}) = {\sqrt{2} E \over D} \left( 1 - {\delta_1^2 \over
4D(\delta_1-2\delta_2)} \right) \left( 1 + {\delta_1^2 - 8\delta_2^2 \over
4D(\delta_1-2\delta_2)} \right)^{-1}, \\
&& Im(\epsilon_{12}) = {\sqrt{2} F \over 3D} \left( 1 - {\delta_1^2 \over
4D(\delta_1-2\delta_2)} \right) \left( 1 - {\delta_1^2 + 8\delta_2^2 \over
12D(\delta_1-2\delta_2)} \right)^{-1}, \\
&& \epsilon_{13} = - {\delta_1 \over 4D} \left( 1 + {\delta_2 \over D}
\right)^{-1}, ~~~ \epsilon_{23} = {\delta_1 \over 2 \sqrt{2} D^2} (E + iF).
\end{eqnarray}
Using Eq.~(26), and neglecting $\alpha'_3$, we obtain
\begin{equation}
\tan \delta_{CP} = {F \over D} \left( 1 + {\delta_2 \over D} \right)
\left[ 1 - {E \over D} \left( 1 + {\delta_2 \over D} \right) \right]^{-1}.
\end{equation}

Assuming inverted hierarchy, we let
\begin{equation}
A = D + \delta_1, ~~~ B = -2D + \delta_2, ~~~ C = E + iF,
\end{equation}
then
\begin{eqnarray}
&& m'_1 = m_1 = 3D^2 + {D \over 2} (\delta_1 - 2 \delta_2), ~~~
m'_2 = m_2 = -3D^2 + D (\delta_1 + 4 \delta_2), \\
&& m'_3 = m_3 = -D^2 + {D \over 2}(\delta_1 + 2 \delta_2), \\
&& m_4 = {D \over 2} \delta_1, ~~~ m_5 = - {\delta_1 \over \sqrt{2}}
(E + iF), ~~~ m_6 = - {E + iF \over \sqrt{2}} (6D + \delta_1 - 2 \delta_2), \\
&& Re(\epsilon_{12}) = -{E \over D \sqrt{2}} \left( 1 + {\delta_1 \over
6D} - {\delta_2 \over 3D} \right) \left( 1 - {\delta_1  \over
12D} - {5\delta_2 \over 6D} \right)^{-1}, \\
&& Im(\epsilon_{12}) = {2 \sqrt{2} F \over \delta_1 + 2 \delta_2}
\left( 1 + {\delta_1 \over
6D} - {\delta_2 \over 3D} \right), \\
&& Re(\epsilon_{13}) = {\delta_1 \over 8D} \left( 1 - {\delta_2 \over 2D}
\right)^{-1}, ~~~ Im(\epsilon_{13}) = - {F \delta_1 \over 16 \sqrt{2} D^2}, 
\end{eqnarray}
and $\epsilon_{23}$ is determined by
\begin{equation}
-\epsilon_{23}^* m_2 + \epsilon_{23} m_3 = -m_5 + \epsilon_{12}^* m_4 +
\epsilon_{13}^* m_6 - \epsilon_{13}^* \epsilon_{12}^* m_1.
\end{equation}

For our numerical analysis, we set
\begin{eqnarray}
&& \Delta m^2_{21} = 7.59 \times 10^{-5}~{\rm eV}^2, ~~~ 
\Delta m^2_{32} = 2.45 \times 10^{-3}~{\rm eV}^2, \\ 
&& \sin^2 2 \theta_{12} = 0.87, ~~~  
\sin^2 2 \theta_{13} = 0.092.  
\end{eqnarray}
We then diagonalize Eq.~(15) exactly and scan for solutions satisfying the 
above experimental inputs. 
Assuming normal hierarchy, we find $\sin^2 2 \theta_{23}$ to range from 
0.9501 for $\delta_{CP}=0$ to 0.9505 for $|\tan \delta_{CP}| = 0.2$, as 
shown in Fig.~1.  
This is an imperceptible change, so our model prediction 
for $\sin^2 2 \theta_{23}$ is basically unchanged from the real case. 
We show the absolute values $|A|$, $|B|$, $|D|$, and $|C|$ as functions 
of $\sin^2 2 \theta_{23}$ in Fig.~2, 
and $E$ versus $F$ in Fig.~3.  
As expected, $F/E$ may be large, but $\tan \delta_{CP} \simeq F/D$ 
remains small.  We then plot the three physical neutrino masses $m'_{1,2,3}$ 
as well as $m_{ee}$ as functions of $|\tan \delta_{CP}|$ in Fig.~4, 
and the Majorana phases $\alpha_{1,2}$ versus $|\tan \delta_{CP}|$ in Fig.~5.
For inverted hierarchy, we show in Figs.~6 to 10 the corresponding plots. 
We note again that $\tan \delta_{CP}$ is small, but now $\alpha_{1,2}$ 
are much larger.  This can be seen from Eq.~(40) versus Eq.~(32).

In conclusion, we have studied how the $T_7$ model of Ref.~\cite{ckmo11R} 
allows $CP$ violation in the neutrino mixing matrix.  There are three real 
parameters $A,B,D$ and one complex parameter $C=E+iF$, from which nine 
physical observables may be derived.  Given the experimental inputs 
$\Delta m^2_{21}$, $\Delta m^2_{32}$, $\sin^2 2 \theta_{12}$, and the 
recently measured  $\sin^2 2 \theta_{13}$, the remaining five observables 
depend on only one variable which we choose to be $\delta_{CP}$. 
Because of the structure of the neutrino mass matrix constrained by 
$T_7$, even if $C$ has a large phase, i.e. $F/E$ is large, $\tan \delta_{CP}$ 
remains small.  For $\sin^2 2 \theta_{13} = 0.092$ and $\sin^2 2 \theta_{12} 
= 0.87$, we find $\sin^2 2 \theta_{23}$ to be essentially fixed at 0.95 
as $|\tan \delta_{CP}|$ changes from 0.0 to 0.2.  The Majorana phases 
$\alpha_{1,2}$ are comparable to $\delta_{CP}$ in magnitude for normal 
hierarchy, but are much larger for inverted hierarchy.

\newpage
\underline{Acknowledgments}: 
The work of H.I. is supported by Grant-in-Aid for Scientific Research, 
No.~23.696, from the Japan Society of Promotion of Science.
The work of S.K. is supported in part by ICTP Grant AC-80. 
The work of E.M. is supported in part by the U.~S.~Department of Energy 
under Grant No.~DE-AC02-06CH11357.

\bibliographystyle{unsrt}

\begin{figure}[htb]
\includegraphics[width=14cm]{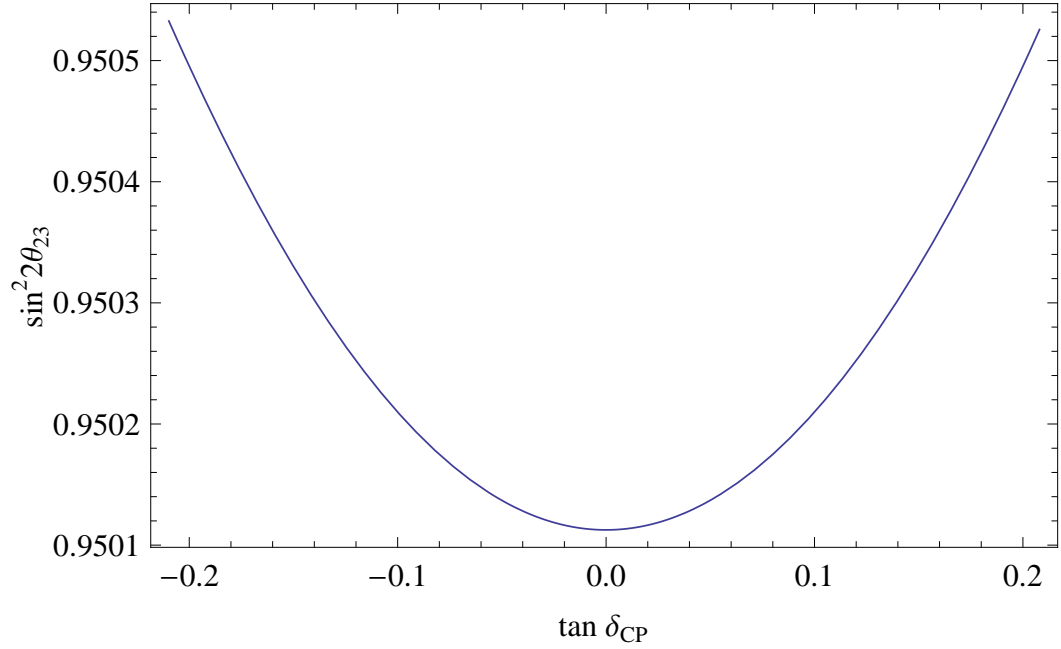}
\caption{$\sin^22\theta_{23}$ versus $\tan \delta_{\rm CP}$ for normal 
hierarchy.}
\end{figure}
\begin{figure}[htb]
\includegraphics[width=14cm]{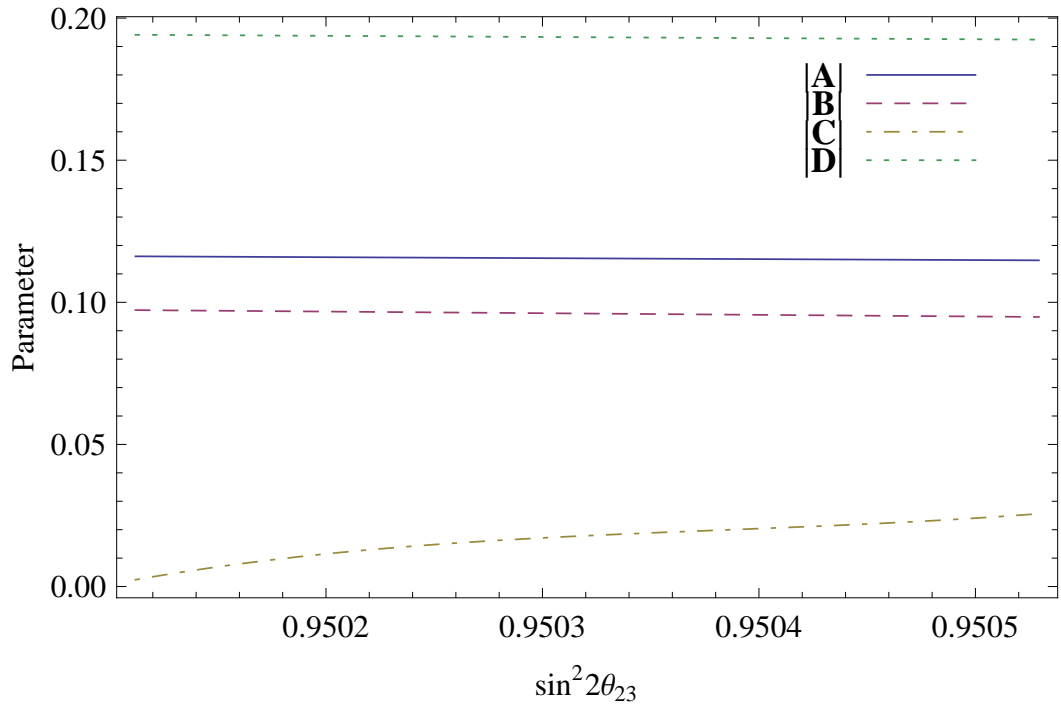}
\caption{$T_7$ parameters for normal hierarchy.}
\end{figure}
\begin{figure}[htb]
\includegraphics[width=14cm]{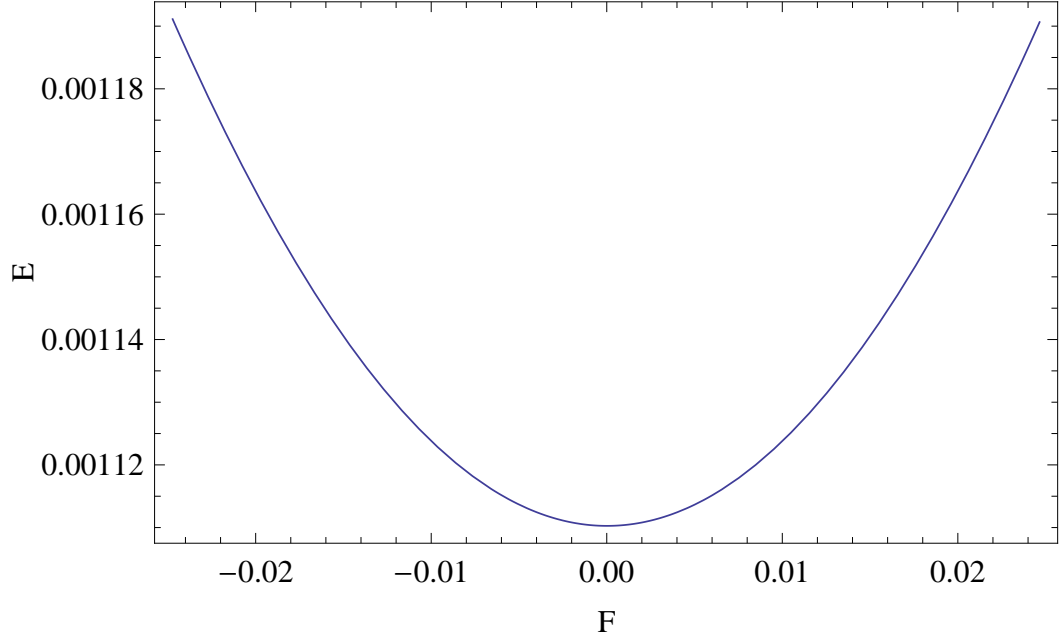}
\caption{$T_7$ complex parameters 
$E$ versus $F$ for normal hierarchy.}
\end{figure}
\begin{figure}[htb]
\includegraphics[width=14cm]{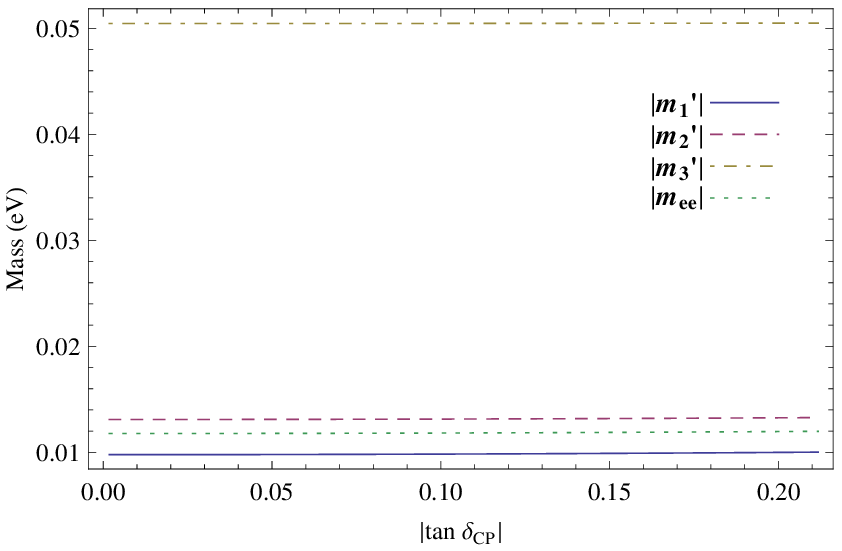}
\caption{Physical neutrino masses and the effective neutrino mass $m_{ee}$ 
in neutrinoless double beta decay for normal hierarchy.}
\end{figure}
\begin{figure}[htb]
\includegraphics[width=14cm]{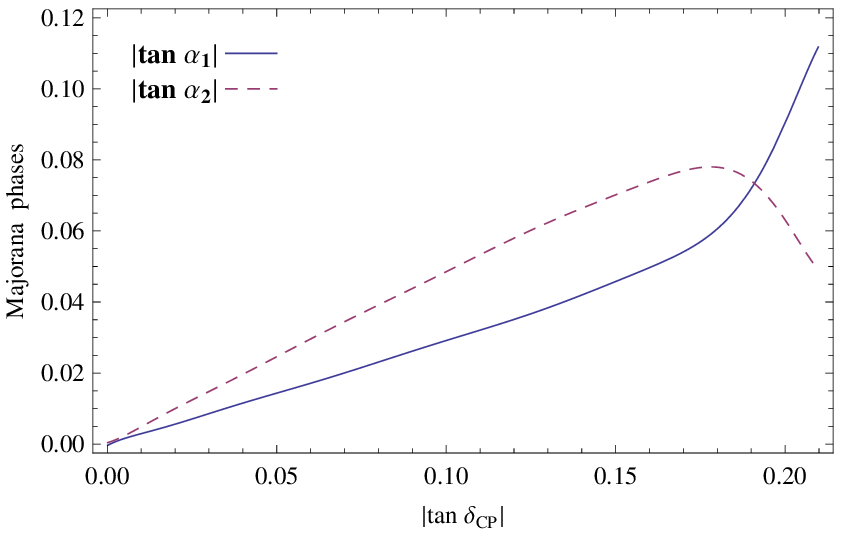}
\caption{Majorana phases 
$|\tan \alpha_1|$ and $|\tan \alpha_2|$  versus 
$|\tan \delta_{\rm CP}|$ for normal hierarchy.}
\end{figure}
\begin{figure}[htb]
\includegraphics[width=14cm]{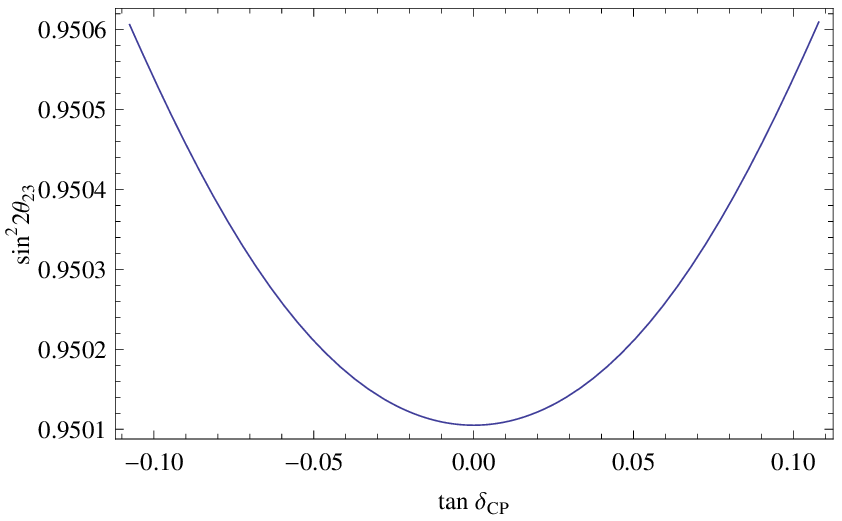}
\caption{$\sin^22\theta_{23}$ versus $\tan \delta_{\rm CP}$ for inverted 
hierarchy.}
\end{figure}
\begin{figure}[htb]
\includegraphics[width=14cm]{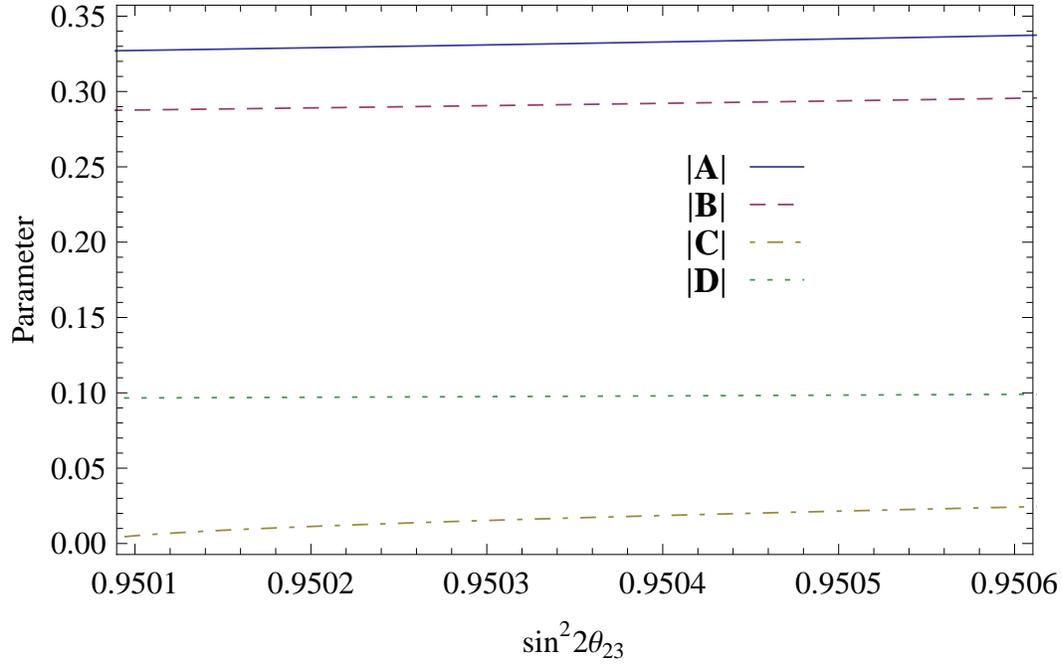}
\caption{$T_7$ parameters for inverted hierarchy.}
\end{figure}
\begin{figure}[htb]
\includegraphics[width=14cm]{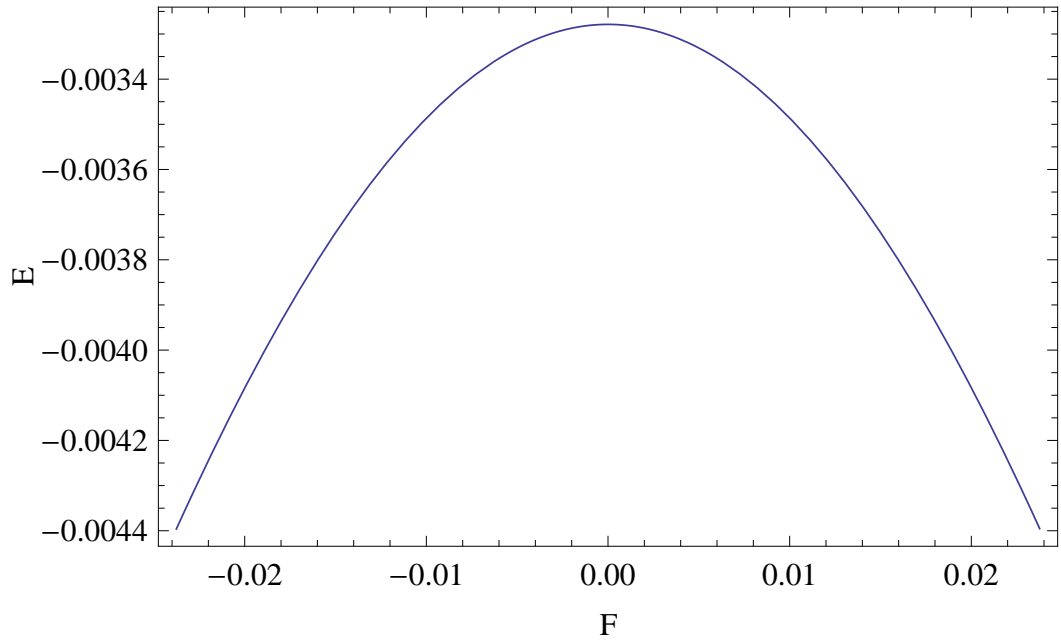}
\caption{$T_7$ complex parameters 
$E$ versus $F$ for inverted hierarchy.}
\end{figure}
\begin{figure}[htb]
\includegraphics[width=14cm]{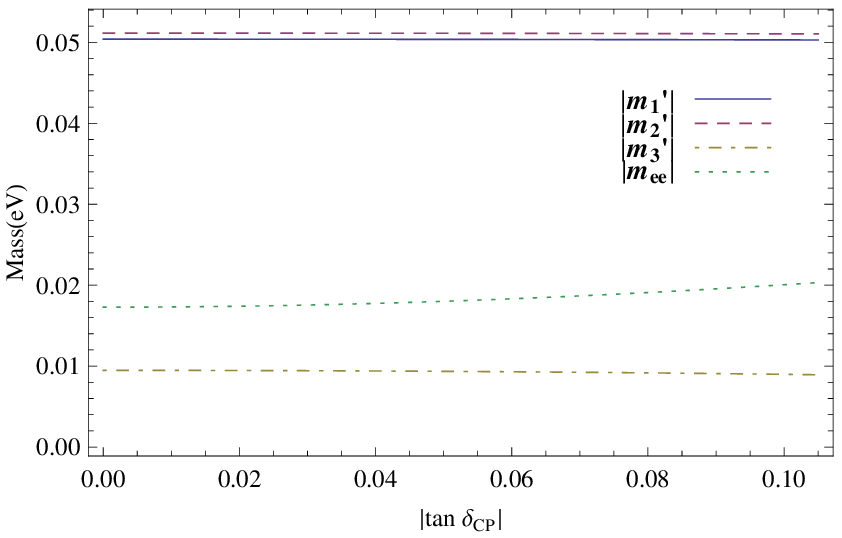}
\caption{Physical neutrino masses and the effective neutrino mass $m_{ee}$ 
in neutrinoless double beta decay for inverted hierarchy.}
\end{figure}
\begin{figure}[htb]
\includegraphics[width=14cm]{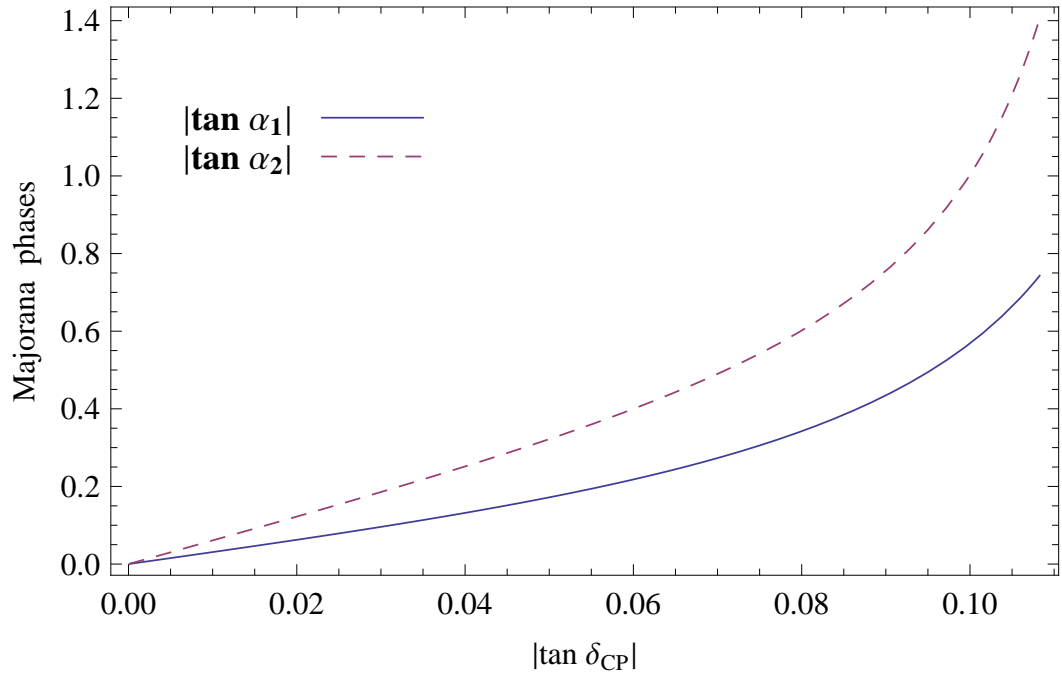}
\caption{Majorana phases 
$|\tan \alpha_1|$ and $|\tan \alpha_2|$  versus 
$|\tan \delta_{\rm CP}|$ for inverted hierarchy.}
\end{figure}

\end{document}